\documentclass[twocolumn,showpacs,preprintnumbers,amsmath,amssymb,natbib,prl]{revtex4}

\usepackage{epsfig}
\usepackage{dcolumn}
\usepackage{bm}

\begin{document}

\title{Full-wave simulations of electromagnetic cloaking structures}

\author{Steven A. Cummer}
\email{cummer@ee.duke.edu}
\author{Bogdan-Ioan Popa}
\author{David Schurig}
\author{David R. Smith}
\affiliation{%
Department of Electrical and Computer Engineering, Duke University, Durham, North Carolina, USA.
}%
\author{John Pendry}
\affiliation{%
Department of Physics, The Blackett Laboratory, Imperial College London, London, UK.
}%

\date{\today}

\begin{abstract}
Based on a coordinate transformation approach, Pendry {\it et al.} have reported electromagnetically anisotropic and inhomogeneous shells that, in theory, completely shield an interior structure of arbitrary size from electromagnetic fields without perturbing the external fields.  We report full-wave simulations of the cylindrical version of this cloaking structure using ideal and nonideal (but physically realizable) electromagnetic parameters in an effort to understand the challenges of realizing such a structure in practice.  The simulations indicate that the performance of the electromagnetic cloaking structure is not especially sensitive to modest permittivity and permeability variations.  This is in contrast to other applications of engineered electromagnetic materials, such as subwavelength focusing using negative refractive index materials.  The cloaking performance degrades smoothly with increasing loss, and effective low-reflection shielding can be achieved with a cylindrical shell composed of an eight (homogeneous) layer approximation of the ideal continuous medium.
\end{abstract}


\maketitle

Pendry {\it et al.} \cite{pendry06} have reported a coordinate transformation approach for designing an electromagnetic material through which electromagnetic fields can be excluded from a penetrable object without perturbing the exterior fields, thereby rendering the interior effectively ``invisible'' to the outside.  Related work has shown how small reflections can be realized through engineered coatings for objects of restricted size and shape \cite{alu05a}, and refractive index profiles have been derived that bend light to produce 2D invisibility based on a conformal mapping approach and assuming the short wavelength geometrical optics limit \cite{leonhardt06}.  The approach in \cite{pendry06} is far more general: it can be applied to problems of any dimension, and it applies under any wavelength condition, not just geometrical optics.  It requires anisotropic media with each permittivity and permeability element independently controlled, but this is within the realm of the metamaterial approach for engineering electromagnetic materials \cite{smith04}.  This approach also requires permittivity and permeability elements with relative magnitudes less than one, and consequently the bandwidth of a passive cloaking material will be limited. 

This cloaking structure shares many qualities with another application based on exotic electromagnetic materials---the negative refractive index perfect lens \cite{pendry99}.  Both are surprising, novel, and of significant theoretical and practical interest.  The physical realizability of each is also not immediately obvious from the original analytical derivation.  The realization of subwavelength focusing is constrained by strict limits on the precise properties of the medium \cite{pendry99,smith03b}, although the effect has been demonstrated in experiment in a number of forms \cite{grbic04, fang05, popa06a}.  Full-wave electromagnetic simulations provided substantial insight into understanding the physical realizability and limitations of this phenomenon.  For example, initial simulations were unable to observe the effect at all \cite{ziolkowski01,loschialpo03}, despite ideal conditions.  Only through great care were later simulations able to demonstrate the effect \cite{cummer03apl1}, which highlighted the delicate conditions needed to produce it and the related difficulty in practically realizing it.

The challenges of realizing the ideal cloaking material are not known at this point.  The analytical derivation does give some clues about its sensitivity to material perturbations.  This structure does not rely on system resonances the way subwavelength focusing does, and the absence of any interfaces between positive and negative permittivity (or permeability) implies no surface resonances should occur.  Additionally, the derivation shows that the transformation required to produce the cloaking material is not unique.  Both of these facts suggest that small perturbations to the cloaking shell properties may not affect the cloaking properties too much.  Whether perfect cloaking is achievable, even in theory, is also an open question.  On the basis of scattering uniqueness, it has been shown that perfect invisibility is not achievable under the Born approximation \cite{wolf93}.  However, the scatterer considered here is not a weak scatterer.  Also, the behavior of the on-axis ray, which cannot be deflected \cite{pendry06}, may degrade cloaking performance to an unknown degree.

In this letter, we report the analysis of full wave simulations with the goal of defining the practically achievable performance of this class of cloaking structure and identifying any challenges in its implementation.  Specifically, we wish to understand the degree to which the on-axis ray and the use of physically realizable, non-ideal materials limit cloaking performance.  The COMSOL Multiphysics finite element-based electromagnetics solver is used for the reported simulations because of the flexibility it allows in specifying material anisotropy and continuous inhomogeneity.

We solve the 2D cylindrical problem in which fields are excluded from an infinite circular cylinder.  Following the approach in \cite{pendry06}, cloaking a central cylindrical region of radius $R_{1}$ by a concentric cylindrical shell of radius $R_{2}$ requires a cloaking shell with the following radius dependent, anisotropic relative permittivity and permeability (specified in cylindrical coordinates):
\begin{eqnarray}
\epsilon_r=\mu_r=\frac{r-R_{1}}{r}, \epsilon_{\phi}=\mu_{\phi}=\frac{r}{r-R_{1}}, \\ \epsilon_z=\mu_z=\left(\frac{R_{2}}{R_{2}-R_{1}}\right)^2\frac{r-R_{1}}{r}.
\end{eqnarray}
The COMSOL solver requires cartesian coordinates, for which the $z$ components do not change but the standard transformation to $x$ and $y$ yields
\begin{eqnarray}
\epsilon_{xx}=\epsilon_r \cos^2\phi+\epsilon_{\phi} \sin^2\phi, \\\epsilon_{xy}=\epsilon_{yx}=(\epsilon_r-\epsilon_{\phi})\sin\phi\cos\phi, \\
\epsilon_{yy}=\epsilon_r \sin^2\phi+\epsilon_{\phi} \cos^2\phi,
\end{eqnarray}
with $\bar{\bar{\mu}}=\bar{\bar{\epsilon}}$ completing the material tensor description.

\begin{figure}[t]
{\epsfig{file=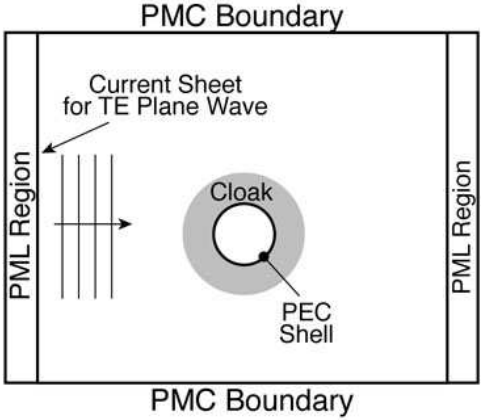,width=2.5in}}
\caption{\label{fig:simdetails} Computational domain and details for the full-wave cloaking simulation.}
\end{figure}

Figure \ref{fig:simdetails} shows the computational domain in which a 2 GHz transverse electric (TE) polarized time harmonic uniform plane wave is incident on a perfect electrical conductor (PEC) shell of diameter 0.2 m (1.33 wavelengths) surrounded by a cloaking shell as specified above with outer diameter 0.4 m (2.67 wavelengths).  Although analytically the inner region is shielded from external fields even without the PEC shell, a penetrable center implies that the internal fields are not unique.  This results in an ill-conditioned problem and, not surprisingly, the simulated solutions are better behaved if a PEC shell is used to exclude the interior from the calculation.  Note that the presence of the PEC shell does not affect in any way the cloaking operation of the system; any object can be present in the interior region, and ideally the external fields are unperturbed whether or not the PEC shell is present.

\begin{figure}[t]
{\epsfig{file=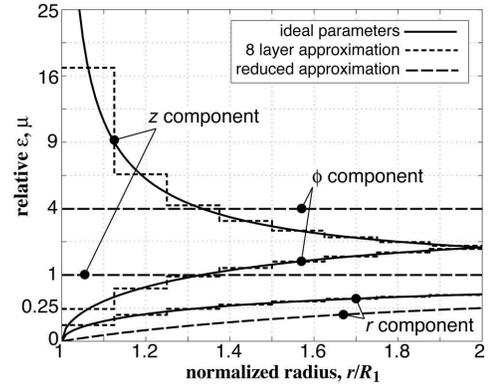,width=2.5in}}
\caption{\label{fig:epsmu} The permittivity and permeability components used in cases 1, 3, and 4 described in the text.}
\end{figure}

Different configurations were simulated to explore the sensitivity and realizability of the cloaking function.  Material parameters (in cylindrical coordinates) for these different configurations are shown in Figure \ref{fig:epsmu}, assuming $R_2=2R_1$.  Case 1 is the ideal cloak material parameters, with lossless $\bar{\bar{\epsilon}}(r)$ and $\bar{\bar{\mu}}(r)$ as defined above.  In case 2, the ideal cloak material parameters are used but with the addition of loss to give a constant electric and magnetic loss tangent of 0.1.  The metamaterial approach in the GHz range can achieve loss tangents significantly lower than this value \cite{greegor03}.  In case 3, the cloaking structure is implemented in a step-wise homogeneous eight layer approximation of the ideal, lossless continuous parameters.  This represents the necessary approximation of realizing a continuous medium with a finite number of discrete layers.

\begin{figure*}
{\epsfig{file=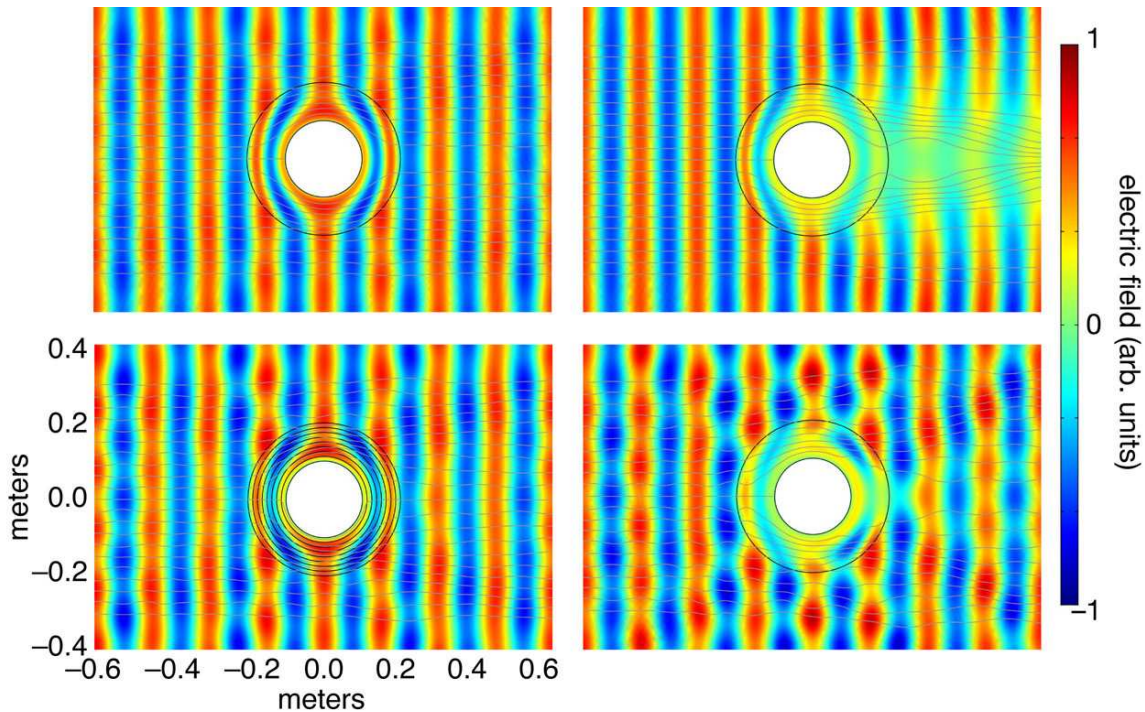,width=6.0in}}
\caption{\label{fig:results} The resulting electric field distribution in the vicinity of the cloaked object.  Power flow lines (in gray) show the smooth deviation of electromagnetic power around the cloaked PEC shell.  In all cases power flow is from left to right.  Upper left (case 1): Ideal parameters.  Upper right (case 2): Ideal parameters with a loss tangent of 0.1.  Lower left (case 3): 8-layer stepwise approximation of the ideal parameters.  Lower left (case 4): Reduced material parameters.}
\end{figure*}

Case 4 is an approximate realization with simplified permittivity and permeability derived as follows.  Limiting ourselves to TE fields and letting $D_z=\epsilon_z E_z$, the Maxwell equations inside the cloaking material are 
\begin{eqnarray}
j\omega D_z=\frac{1}{r}\left[\frac{\partial (r H_\phi)}{\partial r}-\frac{\partial H_r}{\partial \phi}\right] \\
j\omega\mu_r H_r=\frac{1}{r}\frac{\partial (D_z/\epsilon_z)}{\partial \phi} \\
j\omega\mu_{\phi} H_{\phi}=-\frac{\partial (D_z/\epsilon_z)}{\partial r}.
\end{eqnarray}
If $\epsilon_z$ is spatially uniform, then the above equations depend on only two material parameters ($\mu_r \epsilon_z$ and $\mu_\phi \epsilon_z$) instead of the original three.  This gives the ability to choose one of the three arbitrarily to achieve some favorable condition.  One good choice would be to select $\mu_\phi=1$ so that $\epsilon_z=\left(\frac{R_{2}}{R_{2}-R_{1}}\right)^2$ and $\mu_r=\left(\frac{r-R_{1}}{r}\right)^2$.  This has the benefit of making only one component spatially inhomogeneous and also eliminates any infinite values, as shown in Figure \ref{fig:epsmu}.  This reduced medium loses its reflectionless property at interfaces with free space, but it maintains the phase front and power flow bending of the ideal cloaking material.  The simulations below show that this reduced cloaking material can demonstrate the basic physics of the this class of cloaking and may present a simpler path to an experimental demonstration.

Figure \ref{fig:results} shows the resulting simulated electric field distribution and electromagnetic power flow lines for these four cases.  Displayed is the real part of the electric field phasor (equivalent to the time domain fields at the instant of time when the source phase is zero) so that the individual phase fronts are clearly visible.  The cloaking effect is especially clear in the ideal case (upper left panel).  Outside the cloaking shell, the plane wave is almost unaltered, as if no scatterer were present.  Inside the cloaking material, the power flow lines are bent smoothly around the PEC shell, and the phase fronts are also warped in the manner predicted by the original transformation \cite{pendry06}.  The fields are smoothly excluded from the interior region with minimal scattering in any direction.  It is immediately evident that the ideal cloaking material must be highly anisotropic because the wave vector (the normal to the wavefront) and the direction of power flow (the streamlines) are not parallel in the cloaking material.  In many locations they are almost perpendicular.  Anisotropy is required to meet this constraint on the phase and ray directions.

The simulated fields shown in each case were computed with approximately 85,000 elements and 340,000 unknowns, although nearly identical solutions were obtained with a coarser mesh.  In the ideal parameter simulation, there are approximately 32 layers of elements spanning the cloaking material region, which is not especially fine considering the strong variation of the material parameters across this region.  It is interesting to note that the integration of the on-axis ray incident on the center of the cylinder stops because power flow is effectively zero at the inner edge.  

To quantify the cloaking performance of this 2D scenario, we computed the total scattered power per unit length of the cylinder (integrated over all angles) by taking the difference between the solution from simulations with and without the cloak and object.  We then normalized this total scattered power by the incident wave power incident on the diameter of the inner PEC sphere per unit length of the cylinder.  This ratio is the scattering width (the 2D equivalent of radar cross section) of the composite object, averaged over all angles, and normalized to the diameter of the cylinder.  For the simulation shown, this ratio is 0.06.  No doubt this could be pushed even smaller with careful tuning of the simulation parameters; for our purposes, however, it quantitatively shows that the cloaking material is quite effective.  Interestingly, the scattered power is almost isotropically distributed over all angles, an unusual property for an object that is not electrically small.

Loss degrades cloaking performance in a straightforward way.  With electric and magnetic loss tangents of 0.1, the upper right panel of Figure \ref{fig:results} shows that lossy cloaking material is still effective.  Almost no power is scattered in any direction except the forward direction, which is unavoidable because the lossy material absorbs almost all of the forward traveling wave power.  With the loss tangent reduced to only 0.01 (not shown), the effect of loss is almost imperceptible.  No significant differences were observed when loss was introduced as a constant imaginary part instead of a constant loss tangent.  Interestingly, the phase fronts to the left of the object are perturbed less in the lossy case than the lossless case, indicating that the addition of loss can improve backscatter performance (but not the forward scatter) of the cloak.  We speculate that this is due to the following.  

The incident wave power is contained near the on-axis ray requires the most aggressive redirection around the object and is thus the power most likely to be imperfectly handled.  We suggest that small deviations from the ideal medium inherent in a discrete simulation scatter this nearly-on-axis wave power in the almost isotropic scattering pattern observed.  With loss, the cloaking layer absorbs some of scattered power before it exits the cloak.  We thus expect that a lossy cloak would isotropically scatter less than the lossless cloak, at the expense of strong forward scattering.  This is precisely what the simulations show.  Further investigation is merited to determine whether performance of the cloak is limited by this nearly-on-axis power.  

Any physical realization of the ideal continuous cloaking material will require discretization of the material parameters.  The lower left panel of Figure \ref{fig:results} shows the field distribution when the cloak material is approximated by 8 discrete and homogeneous cylindrical layers (see Figure \ref{fig:epsmu}).  The discretized cloaking material, with a normalized isotropic scattering width of 0.215, does not perform as well as the continuous material, as reflected by the more perturbed wavefronts outside the cloaking material.  But most of the incident wave energy is still smoothly bent around the central object.  This simulation shows that systematic perturbations to the ideal material parameters on the order of 10\% do not disrupt the basic cloaking physics.  We speculate that random material property perturbations of the same magnitude probably degrade performance to the same degree.

The lower right panel in Figure \ref{fig:results} shows the field distribution when the cloak material is composed of the reduced material described above, in which the $\mu_\phi$ component is set to unity, $\epsilon_z$ is constant, and only $\mu_r$ is spatially varying.  The significant amplitude variation across the phase fronts indicate that scattering in all directions is significant.  But the bending of the individual phase fronts inside the reduced cloaking material is still present.  In fact, images of the electric field phase for the ideal and reduced cases (not shown) are in close agreement.  This reduced material may be an easier (although not ideal) path to an initial experimental demonstration of this cloaking phenomenon.  In its simplest form it would only require fabricating a material with an inhomogeneous $\mu_r$ component, as $\mu_\phi$ is unity and $\epsilon_z$ is real and greater than one, and thus could be realized with an ordinary dielectric.  Moreover, the very large-valued components required by the ideal cloak are eliminated, and only values on the order of unity and smaller, which are easier to realize in practice, are required in this reduced cloaking material.

\begin{figure}
{\epsfig{file=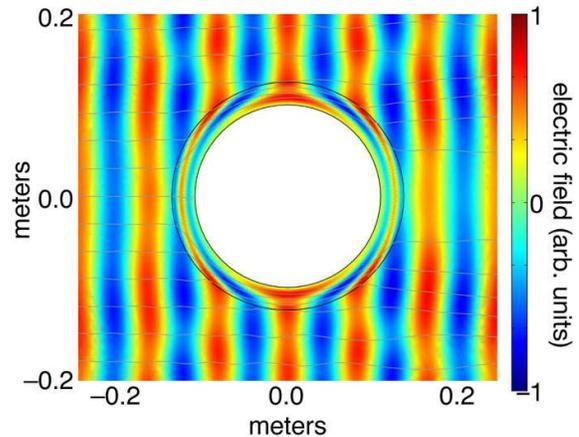,width=3.0in}}
\caption{\label{fig:hero} Simulated cloaking with ideal parameters for a larger cloaked object and thinner cloaking shell.}
\end{figure}

The simulations presented above all contained a cloaking material layer with a thickness parameter $R_{2}/R{1}-1$ (cloak thickness normalized to the radius of the cloaked region) of unity.  One might reasonably ask if the cloaking process becomes substantially more difficult to simulate, and therefore realize, with a thinner cloaking shell and an electrically larger object.  To the extent the computational power is not a limitation, Figure \ref{fig:hero} suggests that the answer is no.  Here we simulate the cloaking of a PEC cylinder of diameter 2.67 wavelengths with a cloaking layer of thickness parameter 0.25 (i.e., the cloak thickness is only 25\% the cloaked cylinder radius).  Even with the more aggressive ray and phase front bending and wavelength compression required to cloak this object, high fidelity simulations of the effect can be achieved with a relatively modest number of elements (approximately 20) across the width of the cloaking shell.  It seems likely that thinner cloaking shells will be require more tightly controlled electromagnetic parameters, but the relative ease with which this effect can be simulated suggests that sensitivity to medium parameters is still modest.  An experimental demonstration of the basic physics of this class of electromagnetic cloaking structure should be possible even with non-ideal electromagnetic metamaterials. 

David Schurig would like to acknowledge support from the IC Postdoctoral Research Fellowship Program.


\end{document}